\documentclass[twocolumn,showpacs,amsmath,aps,pra,amssymb,nofootinbib,superscriptaddress]{revtex4-1} 
\usepackage[T1]{fontenc}
\usepackage[latin9]{inputenc}
\usepackage{times}
\usepackage{color} 
\usepackage{xspace}
\usepackage{amssymb,amsmath}
\usepackage{amsbsy}
   \usepackage{dsfont}
\usepackage{graphicx}
\usepackage{bm}
\usepackage{epstopdf}
\usepackage{float}
\usepackage{adjustbox}

\usepackage{soul}

\usepackage[unicode,breaklinks]{hyperref}
\hypersetup{
    unicode=true,
    a4paper=true,
    plainpages=false, 
    colorlinks=true,
    linkcolor=blue,
    citecolor=blue,
    filecolor=black,
    urlcolor=blue
}
\newcommand{\add}{a_{dd}}
\newcommand{\edd}{\epsilon_{dd}}
\newcommand{\br}{\mathbf{r}}
\newcommand{\bx}{\mathbf{x}}

\newcommand{\LGP}{\mathcal{L}_\mathrm{GP}}
\newcommand{\gammaQF}{\gamma_{\mathrm{QF}}}

\begin{document}
 \title{Excitations of a  vortex line in an elongated dipolar condensate}
\author{Au-Chen Lee}
\affiliation{The Dodd-Walls Centre for Photonic and Quantum Technologies, New Zealand}
\affiliation{Department of Physics, University of Otago, New Zealand}
\author{D.~Baillie}
\affiliation{The Dodd-Walls Centre for Photonic and Quantum Technologies, New Zealand}
\affiliation{Department of Physics, University of Otago, New Zealand}
\author{R.~N.~Bisset}
\affiliation{Institut f\"ur Theoretische Physik, Leibniz Universit\"at Hannover, Germany}
\affiliation{INO-CNR BEC Center and Dipartimento di Fisica, Universit\`a di Trento, Italy}
\author{P.~B.~Blakie}  
\affiliation{The Dodd-Walls Centre for Photonic and Quantum Technologies, New Zealand}
\affiliation{Department of Physics, University of Otago, New Zealand}

\begin{abstract}
We characterise the properties of a vortex line in an elongated dipolar Bose-Einstein condensate.   
 Increasing the strength of the dipole-dipole interactions (DDIs) relative to the short ranged contact interactions we find that the system crosses over to a self-bound vortex droplet stabilized from collapse by quantum fluctuations. 
 We calculate the quasiparticle excitation spectrum of the vortex state, which is important in characterizing the vortex response, and assessing its stability.  When the DDIs are sufficiently strong  we find that the vortex is dynamically unstable to quadrupolar modes.
\end{abstract} 
  
\maketitle
\section{Introduction}
\begin{figure}
\centering
\includegraphics[width=\linewidth,trim=30 70 30 50,clip=true]{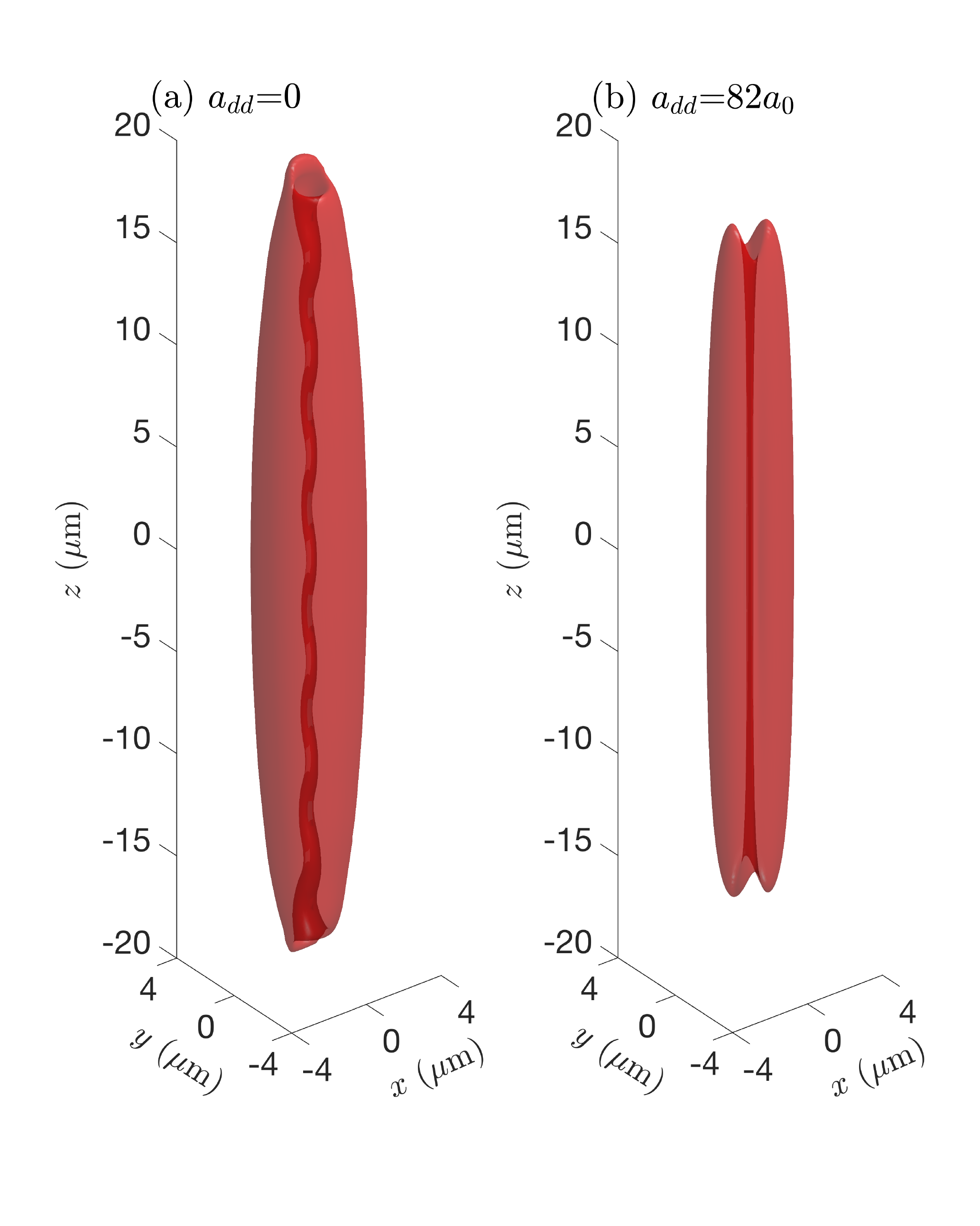}
\caption{ Density isosurface of the $s=1$ vortex state of a condensate for (a) purely contact interactions and  (b) a dipole interaction strength close to instability.  In subplot (a) a Kelvin-wave quasiparticle is superimposed on the condensate causing the vortex line to wiggle [mode (k) indicated in Fig.~\ref{fig:excitations}(a)]. In subplot (b) a  quadrupolar quasiparticle is superimposed on the condensate causing the density around the vortex to split into two pieces [mode (q3) indicated in Fig.~\ref{fig:excitations}(d)].    Isosurfaces indicate a density of $10^{20}\,$m$^{-3}$. 
 }
\label{fig:schematic}
\end{figure}

In this paper we consider the properties of a vortex line in a dipolar condensate [e.g.~see Fig.~\ref{fig:schematic}]. For condensates with short ranged (contact) interactions such a vortex line has been prepared by rotating a cigar shaped trap about its axis of symmetry  \cite{Rosenbusch2002a}. In that system Kelvin waves [e.g.~see Fig.~\ref{fig:schematic}(a)] were observed, emerging from a parametric resonance  with a quadrupolar excitation that could be directly driven with a rotating perturbation \cite{Bretin2003a} (also see \cite{Mizushima2003a,Simula2008b,Simula2008c}).  
To date there has been no reported observation of vortices in a dipolar condensate, however there has been considerable theoretical interest in this topic (e.g.~\cite{Yi2006a,Cooper2005a,Zhang2005a,Komineas2007a,ODell2007a,Klawunn2008a,Wilson2009a,Klawunn2009a,Abad2009a,Bijnen2009a,Mulkerin2013a,Martin2017a}).  
Notably, Klawunn \textit{et al}.~\cite{Klawunn2008a,Klawunn2009a}  found that the DDIs affected the Kelvin modes of a vortex line, and that for negatively tuned DDIs the Kelvin dispersion relation could develop a roton-feature leading to a transverse instability of the vortex line.

The recent observation of quantum droplets formed from a dipolar condensate have opened new directions of research in this system. These droplets occur for sufficiently strong DDIs and arise from the interplay of attractive two-body interactions  and the repulsive quantum fluctuation (QF) effects \cite{Kadau2016a,Ferrier-Barbut2016a,Chomaz2016a,Wachtler2016a,Wachtler2016b,Schmitt2016a,Baillie2016b,Bisset2016a}. 
Irrespective of their confinement, dipolar quantum droplets tend to have an elongated (prolate) density distribution with the long axis in the direction that the dipoles are polarized.  Recently Cidrim \textit{et al.}~\cite{Cidrim2018a} considered whether these droplets might be able to support a vortex. They presented predictions for vortex droplet stationary states, but observed that under time evolution these states were highly unstable with a tendency to split into two parts.

The primary system we consider here is an elongated dipolar condensate confined in a prolate harmonic trap with a vortex line on its long axis. We use extended meanfield theory to calculate stationary vortex states. This theory includes the effects of QFs, within a local density approximation, allowing us to study the system in the condensate and droplet regimes. For the trap geometry we consider, we find that the condensate continuously transforms into a vortex droplet as the DDIs increase in strength relative to the contact interactions, thus demonstrating a viable scheme for producing vortex droplets.

We also solve the Bogoliubov-de Gennes equations for the quasiparticle excitations. This allows us to quantify the effect of the DDIs and QFs on the Kelvin wave modes, and other relevant low energy modes, and to assess the origin of dynamical instabilities in the system. We find that the first strong instabilities to emerge are quadrupolar in character, causing the condensate to break into two pieces [e.g.~see Fig.~\ref{fig:schematic}(b)], consistent with the decay dynamics seen in Ref.~\cite{Cidrim2018a}. By turning off the QF term in the generalized meanfield theory we can assess the effect of this term on stability and the excitation spectrum of the system. Our results show that the QF terms can have marked differences in the spectral properties, even before the system is in droplet regime. Furthermore, comparison of our results to experiments or alternative theories may be useful in establishing the accuracy of the QF term (in the local density approximation) to vortex states.

We briefly outline the paper. In Sec.~\ref{SecFormalism} we present the generalized meanfield theory for the stationary state and the associated formalism for the quasiparticle excitations. The main results are presented in Sec.~\ref{SecResults}. We begin by examining the stationary state properties, and the crossover to the vortex droplet state as the DDIs increases (with the QF term) or the mechanical collapse of the condensate (without the QF term). We then present the related excitation spectrum focusing on the low energy branches and identify the modes that cause the vortex to become dynamical unstable. We then conclude our work.

\section{Formalism}\label{SecFormalism}

\subsection{Generalized meanfield theory}
 The stationary states of a dipolar condensate are described by the generalized Gross-Pitaevskii equation (GPE) (e.g.~see \cite{Saito2016a,Ferrier-Barbut2016a,Wachtler2016a,Schmitt2016a,Wachtler2016b,Bisset2016a,Baillie2016b,Chomaz2016a,Boudjemaa2015a,Boudjemaa2016a,Boudjemaa2017a,Oldziejewski2016a,Macia2016a})
 \begin{align}
\mu{\Psi}=\LGP\Psi,\label{GPE}
 \end{align}
 where
 \begin{align}
    \LGP &\equiv  -\frac{\hbar^2\nabla^2}{2M}  + V_{\mathrm{tr}}+  \Phi(\bx) + \gammaQF |\Psi|^3,\label{e:LGP}
\end{align} 
is the GPE operator, $\Psi$ is the condensate field, and $\mu$ is the chemical potential.
Here we will consider a cylindrically symmetric harmonic trap
$V_{\mathrm{tr}}=\frac{1}{2}M(\omega_\perp^2\rho^2+\omega_z^2z^2)$, 
where $\rho=\sqrt{x^2+y^2}$ is the radial coordinate and  $\{\omega_\perp,\omega_z\}$ are the trap frequencies.  
The effective potential $ \Phi(\bx)\!=\! \int d\bx' \,U(\bx\!-\!\bx')|\Psi(\bx')|^2$ describes the two-body interactions where
\begin{align} 
   U(\br) &= g_s\delta(\br) +\frac{3g_{dd}}{4\pi r^3} (1-3\cos^2 \theta).
\end{align}
Here $g_s=4\pi a_s \hbar^2/M$ is the $s$-wave coupling constant, $a_s$ is the $s$-wave scattering length, and $g_{dd}=4\pi \add \hbar^2/M$ is the DDI coupling constant, with $\add=M\mu_0\mu_m^2/12\pi\hbar^2$ the dipole length determined by the magnetic moment $\mu_m$ of the particles. The DDI term describes dipoles polarized along the $z$ axis by an external field, and $\theta$ is the angle between $\br$ and the $z$ axis. The leading-order QF correction to the chemical potential is $\Delta \mu = \gammaQF n^{3/2}$, which is included in Eq.~\eqref{e:LGP} using the local density approximation $n \!\rightarrow \! |\Psi(\bx)|^2$, with coefficient $\gammaQF\! =\! \frac{32}{3}g_s\sqrt{\frac{a_s^3}{\pi}}(1+\tfrac32 \edd^2)$ 
\cite{Lima2011a,Ferrier-Barbut2016a,Bisset2016a} where $\edd\equiv\add/a_s$.

Here our interest is in axial symmetric stationary states of the form
\begin{align}
\Psi_s(\mathbf{x})=\psi_s(\rho,z)e^{is\phi},\label{e:vortstate}
\end{align}
with $\psi_s$ real where  $\phi=\arctan(y/x)$ is the azimuthal angle. Of primary interest is the singly quantized vortex state $s=1$ that has  $\hbar$ per particle circulation about the $z$-axis. We will also present some results for the ground state case $s=0$.

We solve for the vortex stationary states and the excitations using the Fourier-Bessel type approach introduced by Ronen \textit{et al.}~\cite{Ronen2006a} and adapted to the vortex problem by Wilson \textit{et al.}~\cite{Wilson2009a}. We use a cylindrically-cutoff DDI potential (see Lu \textit{et al.}~\cite{Lu2010a}) to improve the accuracy of the interaction matrix elements.

\subsection{Excitations}\label{sec:excitationsformalism}

The collective excitations of this system are Bogoliubov quasiparticles, which can be obtained as a set of normal modes by linearizing the time-dependent GPE $i\hbar\dot{\Psi}=\LGP\Psi$ about a stationary state. This expansion about the vortex state  (\ref{e:vortstate}) is conveniently taken to be of the form
\begin{align}
    \Psi=e^{i(s\phi-\mu t/\hbar)}&\left[\psi_s + \sum_{m,j}\left( \lambda_{mj} u_{mj} e^{i(m\phi-\epsilon_{mj}t/\hbar)}\right.  \right.  \label{psiexpn}\\
  &\left. \left. \qquad\qquad-\lambda_{mj}^*v_{mj}^*e^{i(-m\phi+\epsilon^*_{mj} t/\hbar)}\right)\right]\!\!,\!\nonumber
\end{align}
(e.g.~see \cite{Morgan1998a,Ronen2006a}),  where $\lambda_{mj}$ is the amplitude of the ${mj}$-mode, and $u_{mj}$ and $v_{mj}$ are the quasiparticle modes with respective energy $\epsilon_{mj}$. 
Here  $m$ is the $z$-component of angular momentum (in units of $\hbar$) of the quasiparticles relative to the condensate, while the remaining radial and axial degrees of freedom are enumerated by the quantum number $j$. 
The cylindrically symmetric amplitudes $\{u_{mj},v_{mj}\}$ satisfy the generalized Bogoliubov-de Gennes equations
 \begin{align}
\begin{pmatrix}  \mathcal{L}_{m+s} + X_m & -X_m \\
 X_m &\! -(\mathcal{L}_{m-s} + X_m)\end{pmatrix}\begin{pmatrix} u_{mj} \\ v_{mj}\end{pmatrix}= \epsilon_{mj} \!\begin{pmatrix} u_{mj} \\ v_{mj}\end{pmatrix},\label{BdGmj}
 \end{align}
 where
  \begin{align} 
 \mathcal{L}_n&=\LGP +\frac{\hbar^2n^2}{2M\rho^2}-\mu,\label{EqLn}\\
  X_mf&= \psi_s e^{-im\phi}\!\int \!d\bx'U(\bx\!-\!\bx') e^{im\phi'}f(\rho',z')\psi_s(\rho',z')\nonumber\\
  &+ \tfrac32 \gammaQF |\psi_s|^3f.
 \end{align}
 Also see Ref.~\cite{Baillie2017a} for a discussion of this excitation formalism applied to ground state (i.e.~$s=0$) droplets. 
Physically acceptable solutions with real eigenvalues $\epsilon_{mj}$  can be chosen to satisfy the normalization condition
\begin{align}
\int d\mathbf{x}\,\left( u_{mj}^2-v_{mj}^2\right)=1,\label{posnorm}
\end{align}
which we refer to as positive norm solutions. Equation (\ref{BdGmj}) also admits (unphysical) negative norm solutions for which the integration in \eqref{posnorm} instead yields a value of $-1$. The Bogoliubov-de Gennes equations possess a symmetry such that a negative norm solution in the $m$-subspace $\{\epsilon_{mj},u_{mj},v_{mj}\}$ corresponds to a positive norm solution in the $-m$-subspace with the transformation: $\epsilon_{mj}\to-\epsilon_{-mj}$, $u_{mj}\to v_{-mj}$ and $v_{mj}\to u_{-mj}$. For complex eigenvalues the excitations can exponentially grow, and the system is dynamically unstable. For this case the integration in \eqref{posnorm} instead yields zero, so it is not possible to construct normalised excitations.

 The numerical solution of Eq.~\eqref{BdGmj} for the case of vortex stationary states is reasonably challenging, and details of our approach will be presented elsewhere \cite{LBB}.
 
\section{Results}\label{SecResults}
\begin{figure}
\centering
\includegraphics[trim=10 0 0 0,clip=true,width=1.0\linewidth]{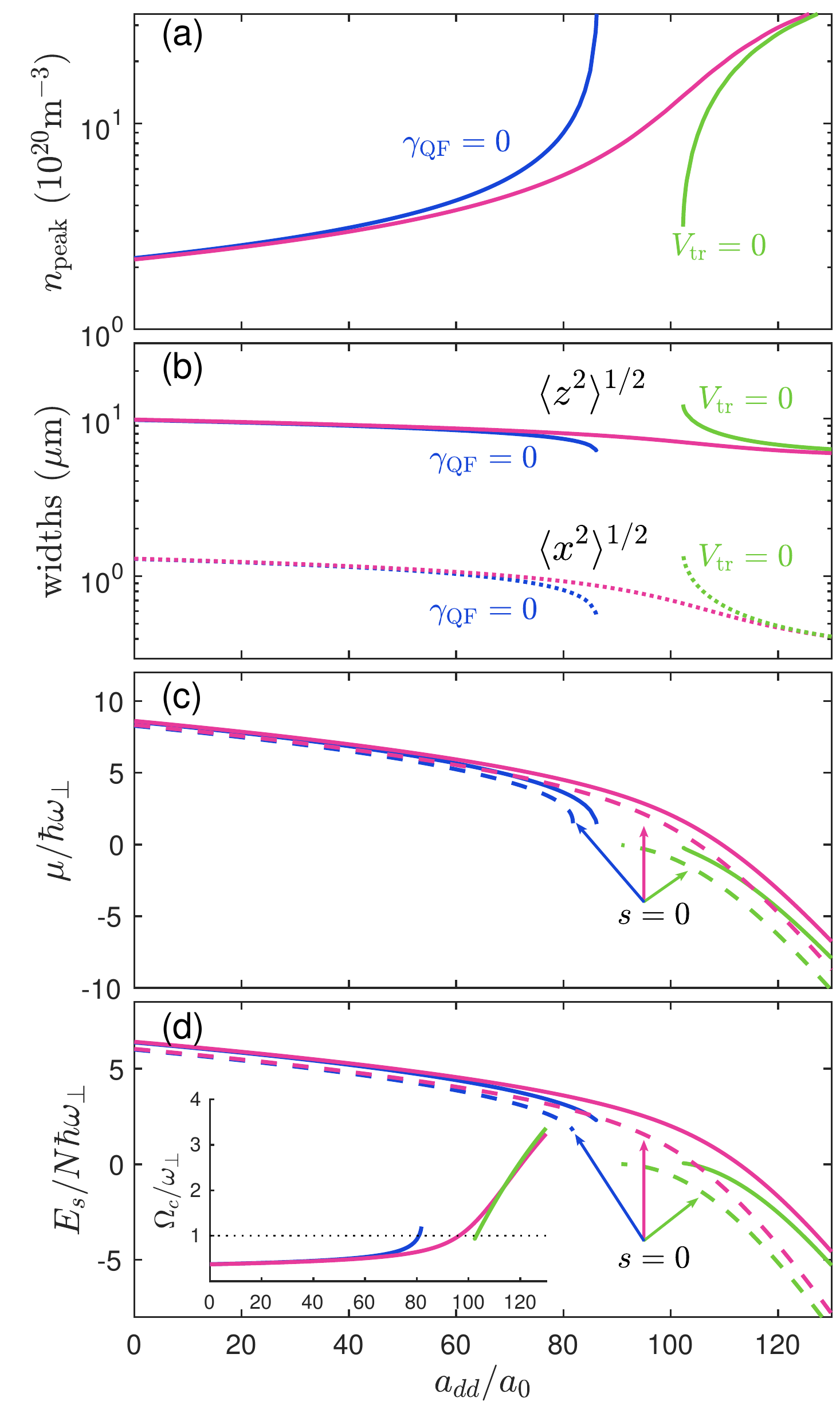}
\caption{Comparison of trapped condensate properties with (magenta lines) and without (blue lines) QF corrections as $a_{dd}$ varies. Free space self-bound droplet solutions (green lines) are also shown. (a) Peak density $n_{\mathrm{peak}}=\max(\psi_s^2)$ of the $s=1$ condensate. (b) Condensate widths given by the rms expectations of the $x$ (dotted) and $z$ (solid line) coordinates. (c) Chemical potential and (d) energy per particle of the $s=0$ ground state (dashed lines) and $s=1$ vortex state (solid lines). Inset to (d): The thermodynamic critical rotation frequency for the $s=1$ vortex state.  The dotted horizontal line indicates the radial trap frequency for reference.
 }
\label{fig:BECproperties}
\end{figure}

\begin{figure*}
\centering
\hspace*{-0.25cm}\includegraphics[trim=10 0 0 0,clip=true,width=1.02\linewidth]{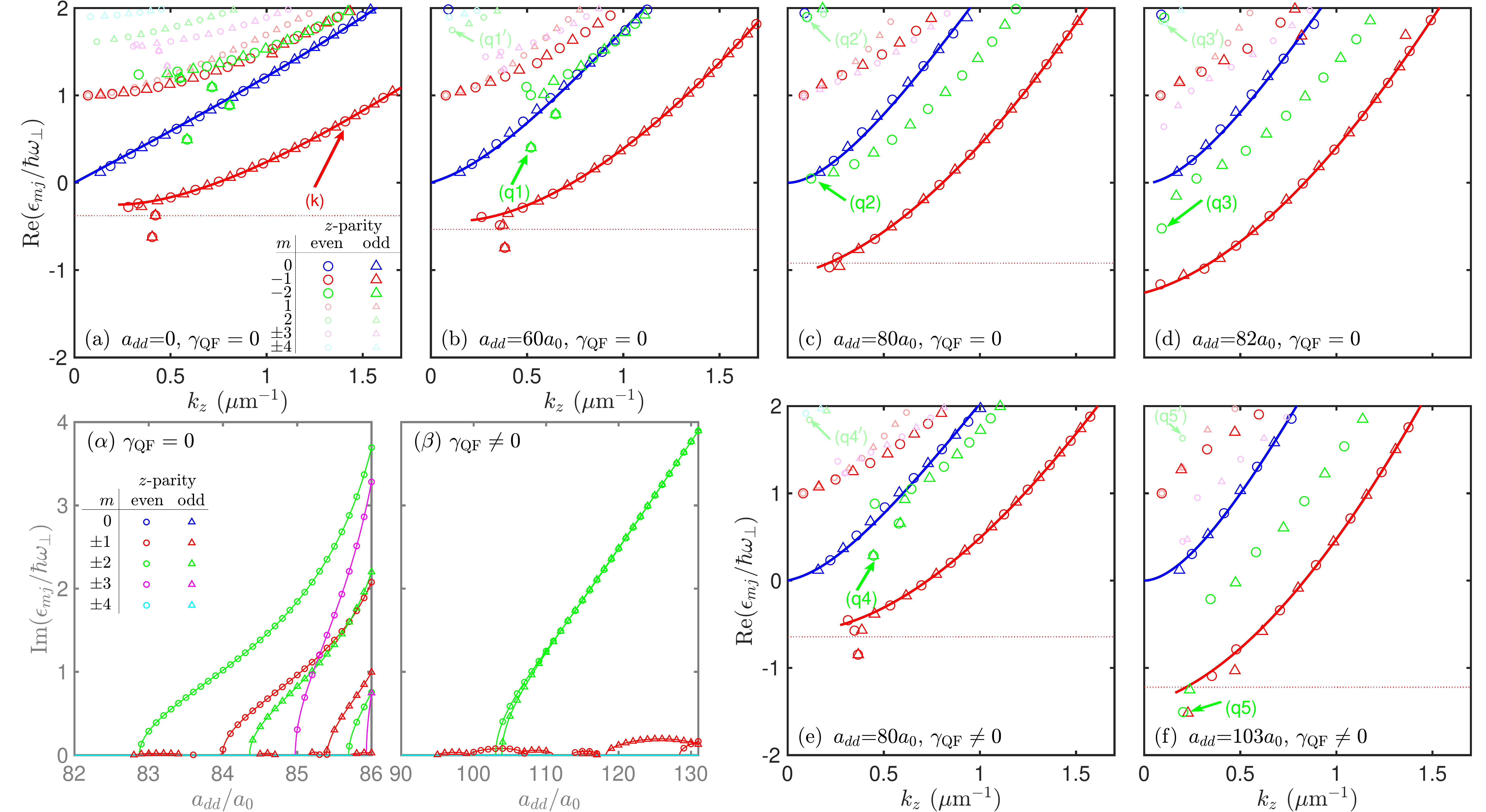}
\caption{(a)-(f) Quasiparticle excitations of an $s=1$ vortex  with $-4\le m\le4$ are shown for various $a_{dd}$ as indicated in each plot.  Subplots (a)-(d) show results without QF corrections, while (e) and (f) include QF corrections. The parity of excitations along $z$ is even (circles) or odd (triangles).
The solid lines are dispersion relation fits (see text) and the horizontal dotted line indicates $-\Omega_c$ [also see inset to Fig.~\ref{fig:BECproperties}(d)]. Subplots ($\alpha$) and ($\beta$) show the imaginary parts of dynamically unstable modes. The labels (k) and (q1) to (q5) identify modes we discuss in the text (also see Figs.~\ref{fig:schematic} and \ref{fig:denflucts}).}
\label{fig:excitations}
\end{figure*}

For our calculations we take $N=112\times10^3$ $^{164}$Dy atoms in a cigar shaped trap with $\omega_\perp\gg\omega_z$, choosing the case $(\omega_\perp,\omega_z)/2\pi=(98.5,11.8)\,$Hz to match the trap used in Ref.~\cite{Bretin2003a}, and taking a scattering length of $a_s=80\,a_0$, where $a_0$ is the Bohr radius.

\subsection{Stationary state properties}
 
We present our results for the condensate properties in Fig.~\ref{fig:BECproperties} as a function of the DDI strength, parameterized by the dipole length $a_{dd}$.  The strength of the DDI can be tuned using a rotating magnetic field \cite{Giovanazzi2002a,Tang2018a} up to the  maximum value (in a static field)  of $a_{dd}=131\,a_0$ for $^{164}$Dy.  

In the absence of the QF term the system becomes mechanically unstable to collapse, where the condensate widths are seen to decrease and the density increases rapidly as $a_{dd}$ increases towards  $a_{dd}\approx85\,a_0$  [e.g.~see  Fig.~\ref{fig:BECproperties}(a), (b)].
Because the DDIs are anisotropic this type of collapse instability is dependent on the  geometry  of the system \cite{Ronen2007a,Koch2008a}. Since our trap arranges the condensate into a prolate shape (which enhances the attractive head-to-tail part of the DDI), collapse occurs soon after the interactions become dipole dominated (i.e.~when $a_{dd}>a_s=80\,a_0$).
  
Including the QF term [see magenta line in Fig.~\ref{fig:BECproperties}] stabilizes the system against mechanical collapse, and the condensate density  grows more slowly as $a_{dd}$ increases. In the regime $a_{dd}>85\,a_0$ (where collapse would occur without the QF term) the system crosses over to a quantum droplet, and then becomes self-bound (i.e.~can maintain itself as a localized structure even in the absence of confinement \cite{Baillie2016b,Schmitt2016a}). We can illustrate this by considering the system chemical potential and energy [Figs.~\ref{fig:BECproperties}(c),(d)], which both become negative for $a_{dd}\gtrsim 120\,a_0$, indicating that the state is self-bound \cite{Baillie2016b,Baillie2017a}.
Here the energy is calculated using the energy functional
 \begin{align}
 E_s=\int d\mathbf{x}\,\Psi_s^*\left[-\frac{\hbar^2\nabla^2}{2M}  + V_{\mathrm{tr}}+  \frac{1}{2}\Phi + \frac{2}{5}\gammaQF |\Psi_s|^3\right]\Psi_s.
 \end{align} 
  We can also compare the trapped solutions to free space self-bound solutions, i.e.~stationary solutions of Eq.~(\ref{GPE}) with $V_{\mathrm{tr}}=0$ \cite{Cidrim2018a,Baillie2016b}. These results are shown as green curves in Fig.~\ref{fig:BECproperties} and confirm that the trapping potential plays a minor role in the stationary state properties for sufficiently large $a_{dd}$ values.

  It is more convenient for experiments to tune $a_s$ using a Feshbach resonance, keeping $a_{dd}$ fixed. We have repeated the type of  stationary state analysis  presented in Fig.~\ref{fig:BECproperties} but fixing  $a_{dd}=131\,a_0$ and varying $a_s$, i.e.~starting from an initial value $a_s>a_{dd}$ and then decreasing $a_s$ to bring the system into the regime of dominant dipole interactions. For this case we find that without the QF term the vortex state is unstable to mechanical collapse at $a_s\lesssim124\,a_0$. With the QF terms the system smoothly crosses over to a vortex droplet attaining a negative chemical potential and energy for $a_s\lesssim90\,a_0$. Similar behavior has been observed in experiments \cite{Chomaz2016a}, where a droplet was prepared in a prolate trap geometry, albeit for a non-vortex ($s=0$) case and by reducing $a_s$.   It is not expected that this behavior will persist in traps with oblate geometries where the droplet state and the condensate do not smoothly connect (see \cite{Blakie2016a,Bisset2016a,Ferrier-Barbut2018a}).

We also show results for the energy $E_0$ of the respective $s=0$ ground states in Fig.~\ref{fig:BECproperties}(d). In a non-rotating reference frame these states have a lower energy compared to the vortex states when they exist. However we note that the ground state for $\gamma_{\mathrm{QF}}=0$ collapses at a lower value of $a_{dd}$ than the vortex state, thus there is a small range of $a_{dd}$ values where $E_1$ can be calculated yet $E_0$ is undefined. The energy difference between the $s=0$ and $s=1$ states relates to the thermodynamic critical angular frequency \cite{Fetter2001a}
 \begin{align}
 \Omega_c=\frac{E_{1}-E_0}{N\hbar},
 \end{align}
where $\Omega_c$ is the rotation frequency about the $z$ axis required for the vortex state to become energetically favorable. Our results for $\Omega_c$ [see inset to Fig.~\ref{fig:BECproperties}(d)] show that $\Omega_c$ increases with $a_{dd}$. This behavior was expected for a prolate dipolar condensate within the hydrodynamic approximation \cite{ODell2007a,Bijnen2009a} (cf.~\cite{Abad2009a})\footnote{We emphasize that $\Omega_c$ is the critical frequency required to make the $s=0$ and $s=1$ states energetically degenerate, and does not mean that the $s=1$ state is necessarily dynamically stable when it is rotated at $\Omega_c$.}. For results including the QF term in the droplet regime the critical rotation frequency can exceed  the radial trap frequency  (i.e.~for $a_{dd}\gtrsim95\,a_0$).  We note that the self-bound result (green) terminates at $\Omega_c \approx \omega_\perp$ by coincidence for this choice of interaction parameters.

\subsection{Excitation spectrum}

In Fig.~\ref{fig:excitations} we present the results for the quasiparticle excitation spectra corresponding to stationary states analyzed in Fig.~\ref{fig:BECproperties} for various $a_{dd}$ values, both with and without the QF term. We restrict our attention to excitations with relative angular momentum quantum number $|m|\le4$, which are the lowest energy excitation branches, with higher angular momentum excitations beginning at energies above the range we consider. Our primary focus is on the $m=0,-1,-2$ branches which we discuss further below.  Subplots (d) and (f) show the excitation spectra for $a_{dd}$ close to dynamical instability (i.e.~where the excitation energies develop imaginary parts) for the cases with $\gamma_{\mathrm{QF}}=0$ and  $\gamma_{\mathrm{QF}}\ne0$, respectively. The imaginary parts of the spectrum are shown in Figs.~\ref{fig:excitations}($\alpha$) and ($\beta$) as a function of $a_{dd}$, revealing that the first dynamically unstable modes develop at $a_{dd}\approx82.5\,a_0$ for $\gamma_{\mathrm{QF}}=0$ and at $a_{dd}\approx94\,a_0$ when we include the QF term.

To visualize the spectra we follow the procedure introduced in Ref.~\cite{Simula2008b}\footnote{The results of Fig.~\ref{fig:excitations}(a) are approximately comparable to Fig.~3 of Ref.~\cite{Simula2008b}, although the larger mass of Dy introduces a scaling of the $k_z$-axis. } to map the excitations on to an effective dispersion relation as a function wavevector $k_z$ along the vortex line. This is done by ascribing an average axial wavevector to each excitation according to 
\begin{align}
\langle k_z\rangle_{mj}\equiv\sqrt{\frac{-\int d\mathbf{x}\,u_{mj}^*\frac{\partial^2}{\partial z^2}u_{mj}}{\int d\mathbf{x}\,|u_{mj}|^2}}.
\end{align}
With this mapping we see that the excitations in Figs.~\ref{fig:excitations}(a)-(f) mostly lie on reasonably smooth curves.  Due to finite size effects of the trapped system, some modes fall below these smooth curves. For example consider the lowest two pairs of $m=-1$ ``bending'' modes in Fig.~\ref{fig:excitations}(a). These modes have been analyzed in detail in prior work (see Ref.~\cite{Simula2008b}), and are surface Kelvin modes that have most of their amplitude near the top and the bottom of the condensate.
 
We can arrive at a simple model for the $m=0$ phonon branch based on the assumption that the condensate and excitations have a Gaussian radial profile of the form $\chi(\bm{\rho})=l_\rho^{-1}\rho e^{-\rho^2/2l_\rho^2+i\phi}/\sqrt{\pi}l_\rho$, which has a maximum at $\rho=l_\rho$. For a system that is uniform in $z$, we obtain the dispersion relation \cite{Baillie2017a,Giovanazzi2004a}
\begin{align} 
    \epsilon_{k_z}\!=\!\sqrt{\epsilon_0^2\!+\!2\epsilon_0n_{\mathrm{peak}}c_f\!\left\{g_s\!-\!g_{dd}F_\chi\left(\!\tfrac{k_zl_\rho}{\sqrt{2}}\!\right)\!+\!c_\mathrm{QF}\gamma_{\mathrm{QF}}n_{\mathrm{peak}}^{1/2}\!\right\}},\label{Eqphonon}
\end{align}
where $\epsilon_0=\hbar^2k_z^2/2M$, $c_\mathrm{QF}=\tfrac{18}{25}\sqrt{\tfrac{2\pi e}{5}}\approx1.33$, 
\begin{align} 
F_\chi(q)=1+\tfrac{3}{2}q^2\left[(q^2+2)^2e^{q^2}\mathrm{Ei}(-q^2)+q^2+3\right],\label{Fchi}
\end{align}
 with $\mathrm{Ei}$ being the exponential integral, and $n_{\mathrm{peak}}$ being the peak density.
This result can be applied to our case taking $l_\rho$ as the radius at which the condensate density is maximum in the $z=0$ plane. We have left $c_f$ as a fit parameter\footnote{For a vortex that is uniform along $z$,  and takes the prescribed Gaussian form radially, we have $c_f = e/4 \approx 0.68$.} that accounts for the spatially varying density along $z$, and for our fits, $c_f$ varies from $0.14$ to $0.38$, which is comparable to a similar factor used in Ref.~\cite{Simula2008b}. In Fig.~\ref{fig:excitations}(d) the phonon dispersion curve provided by Eq.~(\ref{Eqphonon}) starts at finite $k_z$ (just visible near $k_z = 0$) because it is imaginary (dynamically unstable) for smaller $k_z$ values, suggesting that the trapped system is stable in this regime due to finite size effects (i.e.~no phonon mode exists with long enough wavelength to access the instability).

We observe that the phonon spectrum changes appreciably as $a_{dd}$ increases, notably changing from being linear to having curvature and growing more rapidly over the range considered. We note that $\mu$ [see Fig.~\ref{fig:BECproperties}(c)], and hence the speed of sound $c=\sqrt{\mu/M}$, decreases with increasing $a_{dd}$.  The speed of sound corresponds to the slope of the dispersion curves in $k_z\to0$ limit. The fitted phonon dispersion lines (\ref{Eqphonon}) indicate that this slope does decrease with increasing $a_{dd}$, although the first discrete excitation in this branch occurs at a $k_z$ value beyond where the linear behavior holds, i.e.~the curvature in the dispersion is already important. This curvature originates from the momentum dependence of the DDIs in the elongated geometry: excitations with $|k_zl_\rho|<1$ experience an attractive DDI that reduces the value of $\epsilon_{k_z}$, while excitations with $|k_zl_\rho|>1$ experience a repulsive interaction that increases $\epsilon_{k_z}$. This behavior is  described  by the $-g_{dd}F_\chi$ term in Eq.~(\ref{Eqphonon}).

The $m=-1$ excitation branch corresponds to Kelvin waves of the vortex line [e.g.~see Fig.~\ref{fig:schematic}(a)]. 
To fit the Kelvin spectrum we use the dispersion relation introduced by Simula \textit{et al.}~\cite{Simula2008b} (also see \cite{Koens2013a,Fetter2004a})
\begin{align}
\omega(k_0+k_z)=\omega_0+\frac{\hbar k_z^2}{2M}\ln\left(\frac{1}{|r_ck_z|}\right),
\end{align}
valid for $|r_ck_z|\ll1$, where $r_c$ is the so called vortex core parameter. Following Ref.~\cite{Simula2008b} we take $r_c$, $k_0$ and $\omega_0$  as a fitting parameters.
In the case of contact interactions the core parameter was found to be weakly dependent on system parameters, even when the healing length changed appreciably (see \cite{Simula2008b,Simula2008c,Koens2013a}).  For our fits (presented in Fig.~\ref{fig:excitations}), we find that $r_c$ changes significantly to accommodate the stiffening of the Kelvin mode excitation branch as $a_{dd}$ increases. E.g., $r_c$ changes from $0.12\mu$m in Fig.~\ref{fig:excitations}(a) to $0.02\mu$m  in Fig.~\ref{fig:excitations}(d).
This stiffening of the Kelvin mode  behavior was predicted for a vortex line in a uniform dipolar condensate in \cite{Klawunn2008a} and given a simple interpretation: The density core in the vortex line can be viewed as a set of holes that effectively interact with each other via the DDI.  For $a_{dd}>0$ these holes minimize energy in a straight line configuration (i.e.~in an attractive head to tail arrangement). The Kelvin modes cause the vortex line to wiggle [e.g.~see Fig.~\ref{fig:schematic}(a)] incurring an energy cost from the repulsive (side-by-side) component of the DDI, hence causing the Kelvin mode energy to increase with increasing DDI strength.

We find in Figs.~\ref{fig:excitations}($\alpha$) and ($\beta$) that the ($m=-1$) Kelvin mode energies can develop an imaginary part for sufficiently large $a_{dd}$ values. Often the magnitude of this imaginary part  remains small, so that these modes are weakly unstable, and will grow slowly. Also the dynamic instability of these modes oscillates as $a_{dd}$ changes. Similar behavior has been seen in other work considering excitations of  vortices (e.g.~see \cite{Kawaguchi2004a,Wilson2009a,Bisset2015b}), and was found to arise from the coupling of modes that are crossing each other as a parameter is changed  (e.g.~see Fig.~4 of \cite{Lundh2006a}).
This suggests that the Kelvin modes will not strongly grow, but that there is a tendency for the vortex line to wobble.  We note that for the case without the QF term [Fig.~\ref{fig:excitations}($\alpha$)], a pair of Kelvin modes grow to have a large imaginary part for $a_{dd}\gtrsim84\,a_0$, but this occurs well after an $|m|=2$ mode has developed as a strong instability.

\begin{figure*}
\centering
\hspace*{-0.2cm}\includegraphics[width=1.02\linewidth]{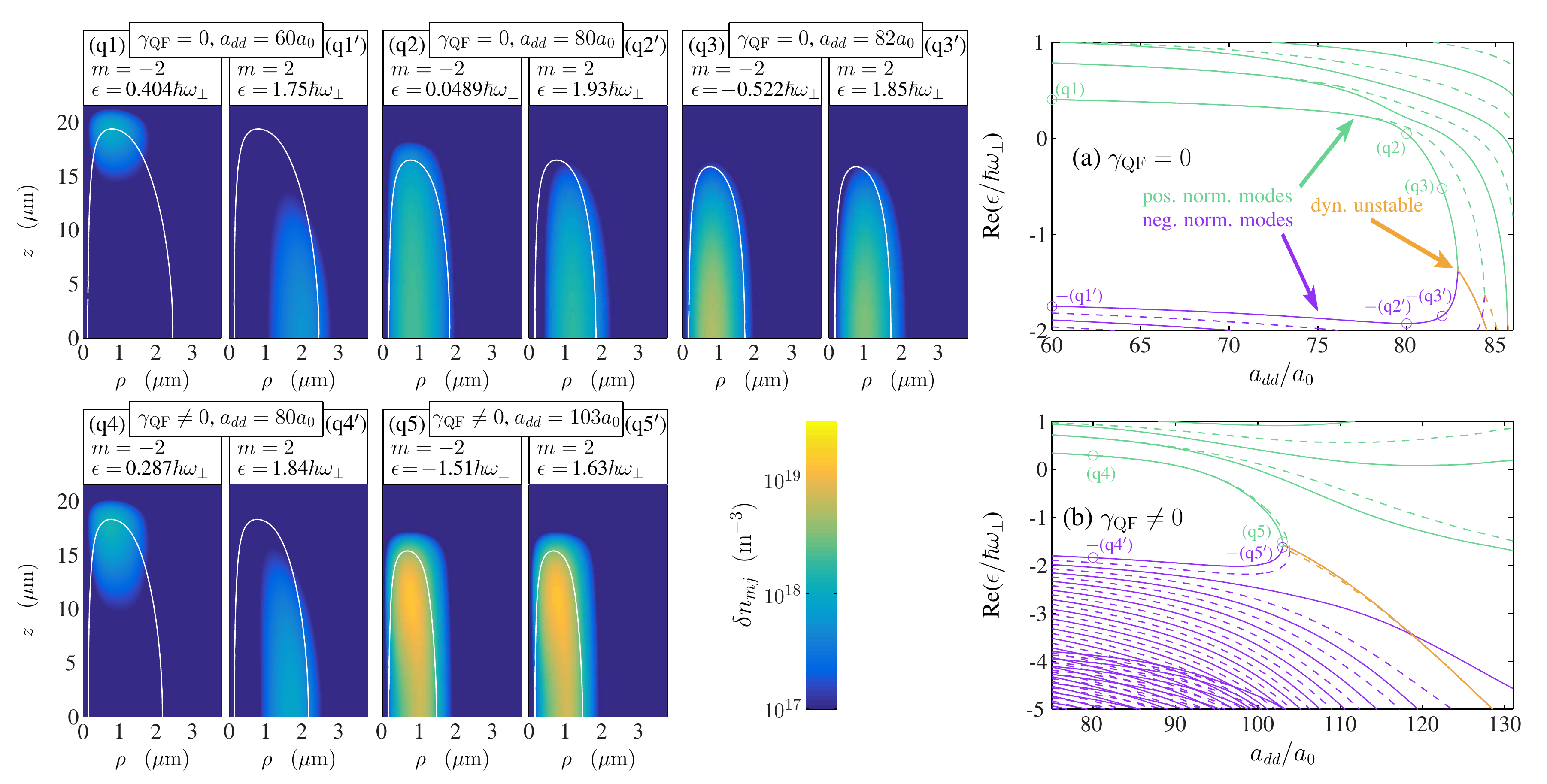}
\caption{ Density fluctuations $\delta n_{mj}$ of various even $z$-parity (q1)-(q5) $m=-2$  and  (q1$^\prime$)-(q5$^\prime$) $m=2$ quadrupolar modes [as labeled in Figs.~\ref{fig:excitations}(b)-(f)].   For reference the white lines indicate a contour of the condensate density at 0.1 of its peak value. Subplots (a) and (b) show the $m=-2$ spectrum, including positive norm (green), negative norm (purple) and the dynamically unstable (orange) modes. Both even (solid line) and odd (dashed line) $z$-parity modes are shown. The (q1)-(q5) mode energies are indicated with small circles. The negative norm $m=-2$ mode energies are the negative of the energies for the corresponding positive norm modes for $m=2$ (see Sec.~\ref{sec:excitationsformalism}). Using this correspondence we also indicate the (q1$^\prime$)-(q5$^\prime$) mode energies on these subplots with small circles.
 }
\label{fig:denflucts}
\end{figure*}

Finally, we consider the $m=-2$ excitation modes, which have a quadrupolar character. As $a_{dd}$ increases, these modes tend to lower their energy relative to the other branches, and notably near instability [see Figs.~\ref{fig:excitations}(d) and (f)]  some of these modes have negative energy.  
Figures~\ref{fig:excitations}($\alpha$) and ($\beta$) reveal that the $|m|=2$ modes are the first to develop large imaginary energies both with and without QFs. This suggests that quite generally the quadrupolar modes will drive the instability of the dipolar vortex line. 

To understand the onset of instability we consider the density perturbation associated with the unstable modes. The density perturbation $\delta n_{mj}$ is the leading order change in the condensate density when we add an $\{mj\}$-quasiparticle to the condensate and is given by 
\begin{align}
\delta n_{mj} = \psi_s(u_{mj}-v_{mj}).
\end{align}
In Fig.~\ref{fig:denflucts} we plot $\delta n_{mj}$ for the lowest energy $m=-2$ mode, which is the first quadrupolar mode to become dynamically unstable. The mode shown is identified as (q1) to (q5) for the different parameter sets and is labeled in Fig.~\ref{fig:excitations} for reference.  Well before instability [i.e.~(q1) for $\gamma_{\mathrm{QF}}=0$ and (q4) for $\gamma_{\mathrm{QF}}\ne0$] the lowest energy quadrupolar mode exists at the surface (top and bottom) of the condensate, and the fluctuation affects the density in these regions. These modes have negligible tunneling through the condensate so that the even and odd $z$-parity modes are degenerate  (see Fig.~\ref{fig:excitations}).  We observe that other degenerate pairs of surface modes often exist, while the rest of the $m=-2$ branch excitations are non-degenerate and fall on a smooth effective dispersion curve.

For the $\gamma_{\mathrm{QF}}=0$ case close to instability [Fig.~\ref{fig:excitations}(c)] the degeneracy is broken between the odd and even modes as the excitation extends through the bulk of the condensate [Fig.~\ref{fig:denflucts}(q2)]. The energy of this mode  descends quickly with increasing $a_{dd}$ as we move closer to instability [Fig.~\ref{fig:excitations}(d)] and the magnitude of the density fluctuation increases significantly  [Fig.~\ref{fig:denflucts}(q3)]. This occurs because the $v$-amplitude changes phase relative to the $u$-amplitude (which also indicates that the excitation is experiencing an effective attractive interaction), thus enhancing $\delta n_{mj}$. In Fig.~\ref{fig:schematic}(b) we indicate the density pattern of the condensate with the (q3) mode coherently added, seeing that this perturbation tends to split the condensate into two parts.

The case with $\gamma_{\mathrm{QF}}\ne0$ progresses towards instability in a similar manner. The degeneracy and hence the top and bottom surface character of the lowest $m=-2$ modes persists to higher values of $a_{dd}$  [see Fig.~\ref{fig:excitations}(e) and Fig.~\ref{fig:denflucts}(q4)], but eventually breaks  when  the surface modes again extend into the bulk  [see Fig.~\ref{fig:excitations}(f) and Fig.~\ref{fig:denflucts}(q5)].

In Fig.~\ref{fig:denflucts}(a) and (b) we see that the dynamic instability occurs when a positive norm and a negative norm quasiparticle mode in the same subspace collide (also see \cite{Kawaguchi2004a,Lundh2006a,Nakamura2008a}).
As we discussed in Sec.~\ref{sec:excitationsformalism}, a negative-norm mode in the $m$-subspace is equivalent to a positive-norm mode in the $-m$-subspace (albeit with an inverted energy sign). Thus the emergence of a dynamically unstable mode in the $m=-2$ subspace will have a partner excitation in the $m=2$ subspace that it will collide with. In subplots (q1$^\prime$) to (q5$^\prime$) of Fig.~\ref{fig:denflucts} we show the $m=2$ excitation that partners with the $m=-2$ mode shown in (q1)-(q5).

It is worth taking a step back to consider the behavior of the quadrupole modes, prior to their instability, in terms of the various energy contributions.
The kinetic energy cost of the azimuthal phase winding differs between the two quasiparticle amplitudes in Eq.~(\ref{psiexpn}), being proportional to $(m+s)^2$ for the  $u_{mj}$ amplitude, and $(m-s)^2$ for the $v_{mj}$ amplitude [see Eqs.~(\ref{BdGmj}) and (\ref{EqLn})].
For $m=-2$ excitations, this places a greater energy cost on the $v_{mj}$ amplitude as compared to  $u_{mj}$.
As a consequence, far before the instability the relevant $m=-2$ excitations are strongly confined to the top and bottom ends of the condensate [Fig.~\ref{fig:denflucts}(q1) and (q4)], minimizing $|v_{mj}|$ by reducing their overlap with the condensate.
In contrast, for the partner $m=2$ excitations [Fig.~\ref{fig:denflucts}(q1$^\prime$) and (q4$^\prime$)] the energy bias is reversed and the energy is reduced by maximizing $|v_{mj}|$, i.e.~the excitation extends throughout the bulk of the condensate.
However, even for $m=2$ the $u_{mj}$ terms still dominates and the density perturbations shown in [Fig.~\ref{fig:denflucts}(q1$^\prime$) and (q4$^\prime$)] clearly exhibit the effects of its larger centrifugal energy, pushing the excitation radially further outwards.
Eventually, for increasing $a_{dd}$ the attractive component of the DDI starts to dominate and the $m=\pm2$ partner excitations begin to hybridize as they approach their instability. As a result, the $m=-2$ excitations overcome their high-density aversion and extend into the bulk of the condensate. The increased tunnelling between the two ends destroys the energetic degeneracy of the odd and even $z$-parity modes.

\section{Conclusions and outlook}
In this paper we have explored the properties of a vortex line in an elongated dipolar Bose-Einstein condensate.   We have presented results for the system properties  as the DDI strength is changed, observing that the system smoothly evolves from being a trap bound vortex into a self-bound vortex droplet as the strength of the DDI  interaction increases. We have also presented results for the quasiparticle excitation spectrum of the system, revealing the behavior of the Kelvin wave and other low energy excitations. In the regime of dominant DDIs we find that this system becomes dynamically unstable to quadrupolar excitations, which appears to be consistent with the decay dynamics observed in GPE simulations of vortex droplets \cite{Cidrim2018a}. More generally, our work suggests that vortices in dipolar droplets are unstable (i.e.~have a short lifetime), cf.~vortices in binary mixture droplets \cite{Kartashov2018a}.

We have presented our results both with and without QFs to reveal their effect on the system. Of course the QFs are necessary for droplet formation at high values of the DDI, but also we observe differences even before this regime [for example in the $m=-2$ excitation modes at $a_s=80\,a_0$, compare Fig.~\ref{fig:excitations} (c),(e)]. Such excitations might be accessible to direct driving (e.g.~see \cite{Bretin2003a}) or could be probed with Bragg spectroscopy using light fields that carry angular momentum (cf.~\cite{Ryu2007a,Bismut2012a,Blakie2012a}). This kind of study would also be useful for gaining a better understanding of the accuracy of the QF treatment we use here which is based on the  local density approximation. 

Experiments have yet to report the observation of vortices in a dipolar condensate. Increased understanding of this system and the regimes where dynamic instabilities occur will be important in future experimental studies.

\begin{acknowledgments}
We thank Tapio Simula and Ryan Wilson  for useful discussions. 
We acknowledge the contribution of NZ eScience Infrastructure (NeSI) high-performance computing facilities, and support from the Marsden Fund of the Royal Society of New Zealand.
RNB was supported by the European Union's Horizon 2020 research and innovation programme under the Marie Sk{\l}odowska-Curie grant agreement No.~793504 (DDQF), the DFG/FWF (FOR 2247), and the Provincia Autonoma di Trento.

\end{acknowledgments}


%

\end{document}